\documentclass[a4paper,11pt]{article}
\usepackage{pos,ulem}
\pdfoutput=1

\makeatletter
     {\bgroup\raggedright\small\section*{\refname
        \@mkboth{\MakeUppercase\refname}{\MakeUppercase\refname}}%
      \list{\name{bib\@arabic\c@enumiv}% HOPE!
            \@biblabel{\@arabic\c@enumiv}}%
           {\settowidth\labelwidth{\@biblabel{#1}}%
            \setlength\itemsep{0pt}
            \setlength\parskip{0pt}
            \setlength\parsep{0pt}
            \setlength\partopsep{0pt}
            \leftmargin\labelwidth
            \advance\leftmargin\labelsep
            \@openbib@code
            \usecounter{enumiv}%
            \let\p@enumiv\@empty
            \renewcommand\theenumiv{\@arabic\c@enumiv}}%
      \sloppy\clubpenalty4000\widowpenalty4000%
      \sfcode`\.\@m}
     {\def\@noitemerr
       {\@latex@warning{Empty `thebibliography' environment}}%
      \endlist\egroup}
\makeatother

\definecolor{orange}{rgb}{1.0, 0.5, 0}
\definecolor{darkgreen}{rgb}{0, 0.6, 0.6}

\title{Composite Higgs scenario in mass-split models}
\author*[a,b]{Oliver Witzel}
\author[a]{Anna Hasenfratz}
\author[a]{Curtis T.~Peterson}
\affiliation[a]{Department of Physics, University of Colorado, Boulder, CO 80309, United States}

\affiliation[b]{Theoretische Physik 1, Naturwissenschaftlich-Technische Fakultät,
Universität Siegen, 57068 Siegen, Germany}

\emailAdd{oliver.witzel@uni-siegen.de}
\emailAdd{curtis.peterson@colorado.edu}

\abstract{Mass-split composite Higgs models naturally accommodate the experimental observation of a light 125 GeV Higgs boson and predict a large scale separation to other heavier resonances. We explore the SU(3) gauge system with four light (massless) and six heavy (massive) flavors by performing  numerical simulations. Since the underlying system with degenerate and massless ten flavors appears to be infrared conformal, this system inherits conformal hyperscaling and allows to study near-conformal dynamics. Carrying out nonperturbative lattice field theory simulations, we present the low-lying particle spectrum. We demonstrate hyperscaling, predict the anomalous mass dimension of the corresponding conformal fixed point, and show that in the investigated mass regime the data are described by dilaton chiral perturbation theory. The proximity of a conformal infrared fixed point leads to a highly predictive particle spectrum which is quite distinct from QCD. Further we present initial results of our finite temperature investigations. \\[3mm] \textbf{For the Lattice Strong Dynamics collaboration}}

\FullConference{%
  40th International Conference on High Energy physics - ICHEP2020\\
  July 28 - August 6, 2020\\
  Prague, Czech Republic (virtual meeting)
}

\begin{document}
\maketitle

\section{Introduction}

The standard model (SM) very successfully describes the interactions of the electro-weak and strong forces. It is however an effective theory and new physics is needed to explain e.g.~dark matter or UV complete the Higgs sector. Focusing at the Higgs sector, experiments have revealed a light, 125 GeV Higgs boson \cite{Aad:2012tfa,Chatrchyan:2012ufa,Aad:2015zhl} but so far no other heavier resonances. This implies that beyond the standard model (BSM) theories aiming to explain the Higgs sector are required to exhibit a large separation of scales \cite{Contino:2010rs,Luty:2004ye,Dietrich:2006cm,Luty:2008vs,Brower:2015owo,Csaki:2015hcd,Arkani-Hamed:2016kpz,Witzel:2019jbe} to match experimental observations.  Such a large separation of scales occurs e.g.~in near-conformal systems with a ``walking'' gauge coupling \cite{Yamawaki:1985zg,Bando:1987br}. To explore the dynamics of near-conformal systems, we use the construction of a mass-split model built on a conformal infrared fixed point (IRFP) \cite{Brower:2014dfa,Brower:2015owo,Hasenfratz:2016gut}. Specifically we study an SU(3) gauge system with $N_\ell=4$ light flavors of mass $\widehat m_\ell$ and $N_h=6$ heavy flavors of mass $\widehat m_h$ \cite{Witzel:2018gxm,Appelquist:2020xua} using the same lattice actions as our investigation of the massless $N_f=10$ system \cite{Hasenfratz:2017qyr,Hasenfratz:2020ess}.

A system with ten massless flavors is conformal i.e.~the gauge coupling is irrelevant and runs to the IRFP. By raising the mass of the heavy flavors, we create a mass-split system that is governed by the nearby IRFP at high energies. The heavy flavors decouple in the infrared where chiral symmetry for the light flavors breaks spontaneously and the gauge coupling starts running again. The heavy flavors play an active role in this setup. Their mass controls the separation between UV and IR scales and effectively sets the hadronic energy scale \cite{Hasenfratz:2017hdd}. The properties of the low energy system differ significantly from a QCD-like setup. Most notably, the isosinglet scalar ($0^{++}$) appears to be light and may need to enter the effective chiral Lagrangian which requires an extension named  dilaton chiral perturbation theory (dChPT) \cite{Golterman:2016lsd,Appelquist:2017vyy,Appelquist:2017wcg,Golterman:2018mfm,Appelquist:2019lgk,Golterman:2020tdq}. The particle spectrum of mass-split models is characterized by the inherited conformal hyperscaling.

In this report we first show conformal hyperscaling in our 4+6 mass-split system and extend/demonstrate dChPT to mass-split systems. In Section \ref{Sec.finiteT} we give a brief outlook on our studies of the underlying finite temperature phase structure. We aim to explore whether this new strongly coupled sector extending the SM could give rise to primordial gravitational waves in the early universe.

\section{Hyperscaling and Dilaton Chiral Perturbation Theory}\label{Sec.Hyper_dChPT}

Properties of mass-split systems can be deduced using arguments based on Wilsonian renormalization group (RG)  methods. We assume that in the UV both mass parameters are much lighter than the cutoff $\Lambda_\text{cut}=1/a$, i.e.~$\widehat m_l \ll 1$, $\widehat m_h\ll 1$ where $a$ denotes the lattice spacing and $\widehat m_{\ell,h}$ are masses in lattice units. Once the energy scale $\mu$ is lowered from the cutoff, the RG flowed lattice action moves in the infinite parameter action space. Both masses increase according to their scaling dimension $y_m$, $\widehat m_{\ell,h} \to \widehat m_{\ell,h} (a \mu)^{-y_m}$. They remain however sufficiently small such that the system stays close to the conformal critical surface and gauge couplings run toward the IRFP.

For this scenario we can use standard hyperscaling arguments \cite{DeGrand:2009mt,DelDebbio:2010ze,DelDebbio:2010jy} to show that any physical quantity $aM_H$ of mass dimension one follows, at leading order, the scaling form \cite{Hasenfratz:2016gut}
 \begin{align}
 a M_H = \widehat m_h^{1/y_m}  \Phi_H( \widehat m_\ell/ \widehat m_h)\,,
\label{eq:M_scaling}
 \end{align}
where $y_m=1+\gamma_m^\star$ is the universal scaling dimension of the mass at the IRFP and $\Phi_H$  some function of $ \widehat m_\ell/ \widehat m_h$. If we consider a ratio of two quantities, Eq.~(\ref{eq:M_scaling}) implies that this ratio will only depend on $\widehat m_\ell/\widehat m_h$.

\begin{figure}[tb]
  \includegraphics[height=0.175\textheight]{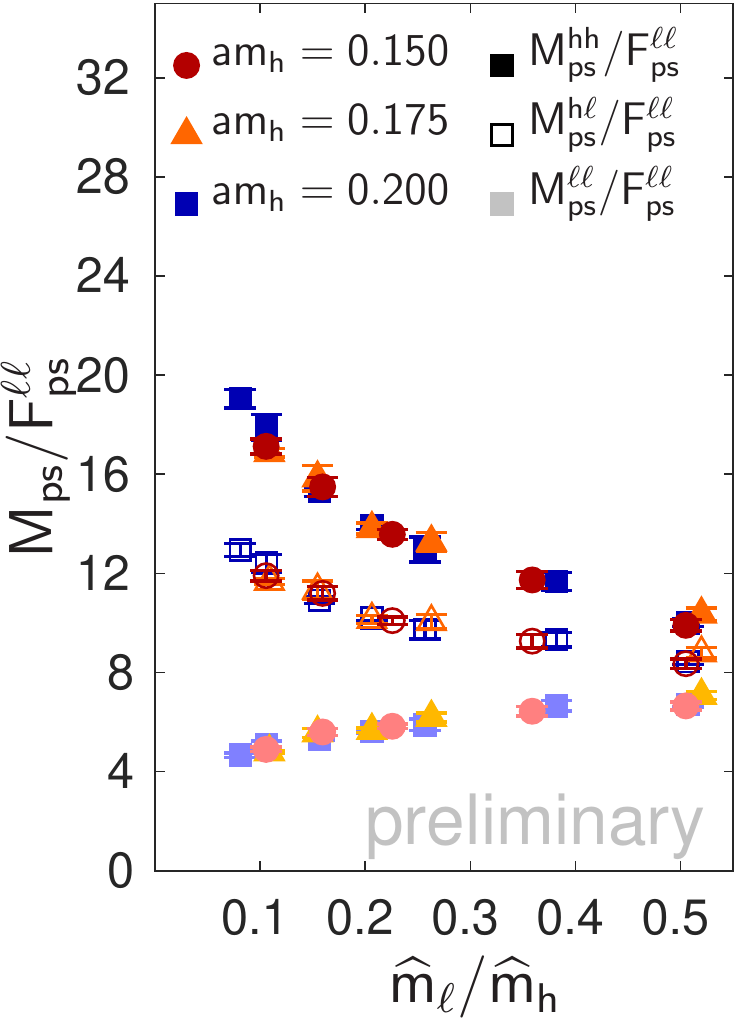}
  \includegraphics[height=0.175\textheight]{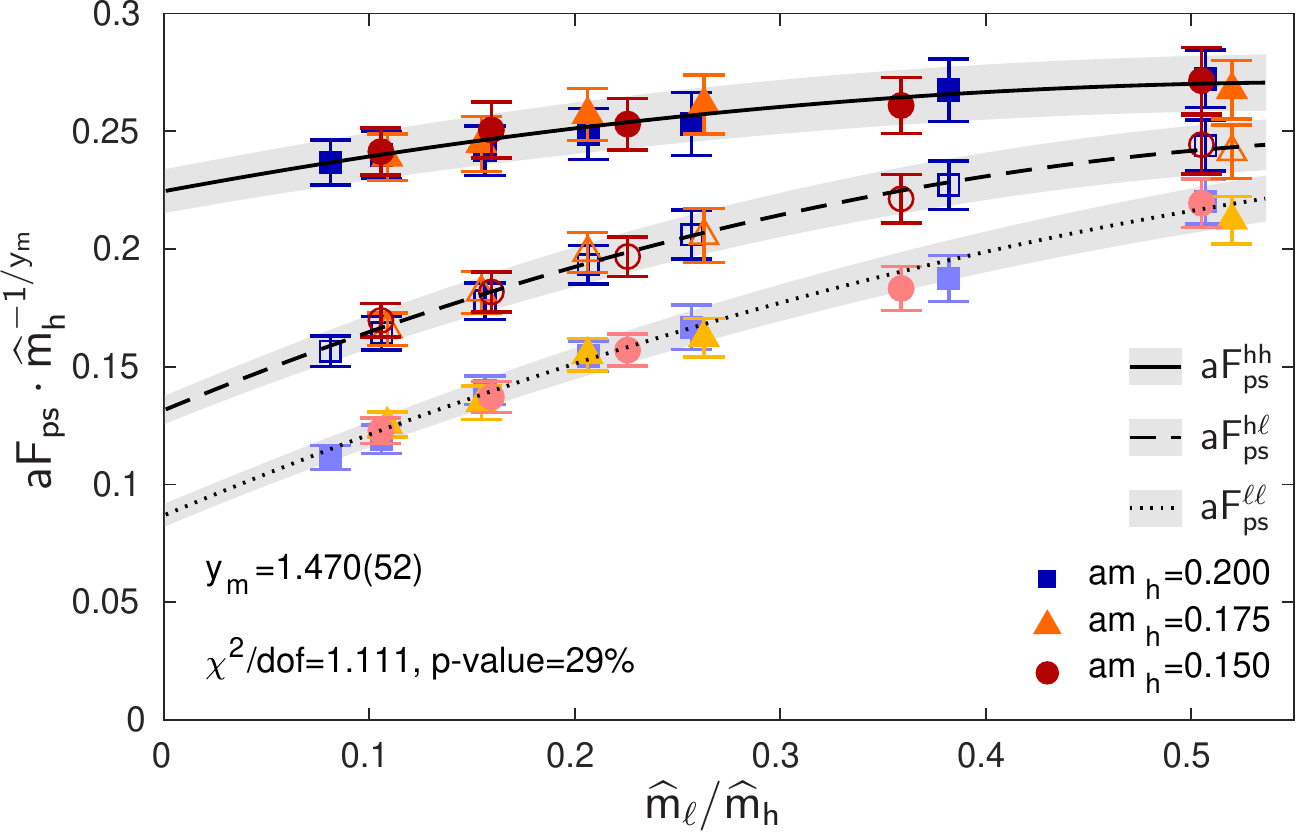}
  \includegraphics[height=0.175\textheight]{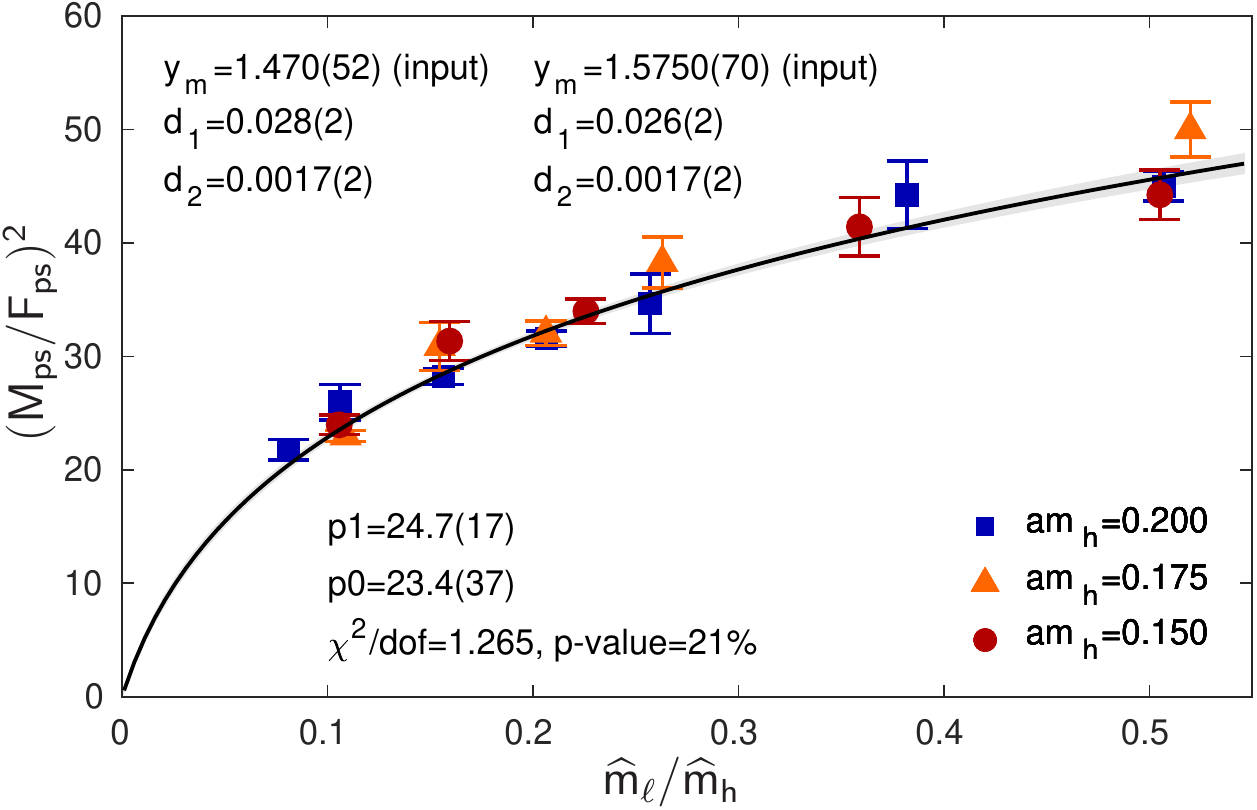}
  \caption{Left: ratio of pseudoscalar mass ($M_\text{ps}$) over pseudoscalar decay constant ($F_\text{ps}$) demonstrating hyperscaling for dimensionless ratios. Middle: combined and correlated fit using the hyperscaling relation (Eq.~(\ref{eq:M_scaling})) to extract the universal scaling dimension $y_m$ from measured $F_\text{ps}$  values. Right: testing dChPT in the light sector assuming a specific form of the dilaton potential.}
  \label{Fig.Hyper_dChPT}
\end{figure}

The left panel of Fig.~\ref{Fig.Hyper_dChPT} demonstrates this by showing ratios of pseudoscalar masses with two heavy, two light, or one heavy and one light flavor in the valence sector over the light-light pseudoscalar decay constant. Despite changing $m_h$ from 0.200 to 0.150, data for the three different ratios (open, filled, and shaded symbols) trace out unique curves.  The high quality of our data even allows us to directly exploit the hyperscaling relation in Eq.~(\ref{eq:M_scaling}) as we show in the central panel. Replacing the unknown functions $\Phi_H$ by a polynomial of second degree, we perform a combined and correlated fit to 51 data points for the pseudoscalar decay constants. The fit exhibits an excellent $p$-value and finds the universal scaling dimension $y_m=1.470(42)$. Since $\gamma_m^\star$ is small, this suggests that $N_f=10$ is sufficiently far from the onset of the conformal window \cite{Yamawaki:1985zg,Matsuzaki:2013eva}. Our value lies in between values predicted for $N_f=8$ and $12$ \cite{Appelquist:2011dp,DeGrand:2011cu,Cheng:2013eu,Cheng:2013xha,Lombardo:2014pda,Ryttov:2016asb,Ryttov:2017kmx,Ryttov:2017lkz,Li:2020bnb}.

Next we consider the low energy infrared limit of our system. Since chiral symmetry of the light sector breaks spontaneously, we expect it can be described by a chiral effective Lagrangian smoothly connecting to the hyperscaling relation Eq.~(\ref{eq:M_scaling}) which is valid at the hadronic scale $\mu=\Lambda_H$. First we express the lattice scale $a$ in terms of the hadronic scale $\Lambda_H$
\begin{align}
   M_H/\Lambda_H    = (aM_H)\cdot{\widehat m}_h^{-1/y_m} &=   \Phi_H( \widehat m_\ell/ \widehat m_h)\,,
\label{eq:M_scaling_LH}
\end{align}
and deduce the relation
 \begin{align}
m_f \propto \widehat m_\ell (a \Lambda_H)^{-y_m}\cdot \Lambda_H  = ( \widehat m_\ell/ \widehat m_h) \cdot \Lambda_H.
\label{eq:m_f}
\end{align}
Equation (\ref{eq:M_scaling}) and the predicted $y_m=1.470(42)$ scaling exponent indicate that our data do not follow the standard chiral perturbative form. We test an extension,  dChPT \cite{Golterman:2016lsd,Appelquist:2017vyy,Appelquist:2017wcg,Golterman:2018mfm,Appelquist:2019lgk,Golterman:2020tdq}, that assumes the presence of a light scalar state. Following \cite{Golterman:2020tdq} we make a specific assumption for the Higgs potential  which leads to the relation
 \begin{align}
&\frac{M_{ps}^2}{F_{ps}^2} = \frac{1}{y_m d_1} W_0\left(\frac{y_m d_1}{d_2} m_f\right),
\label{eq:dChPT2}
 \end{align}
where $W_0$ is the Lambert W-function and $d_1$, $d_2$ are mass independent constants. Using our data from the light sector, we can fit for $d_1$ and $d_2$ as shown in the right panel of Fig.~\ref{Fig.Hyper_dChPT}. This further demonstrates the beauty of our mass-split model. The relevant parameter is $\widehat m_\ell/\widehat m_h$, which is a continuous variable. Since this fit works well, it can be interpreted that our simulations performed so far fall into the range of dChPT. Additional simulations at smaller $\widehat m_\ell/\widehat m_h$ may reveal the need for higher order terms or the applicability of regular chiral perturbation theory.

\section{Investigations of the Finite- and Zero-Temperature Phase Structure}\label{Sec.finiteT}

\begin{figure}
\centering
\includegraphics[width=.305\textwidth]{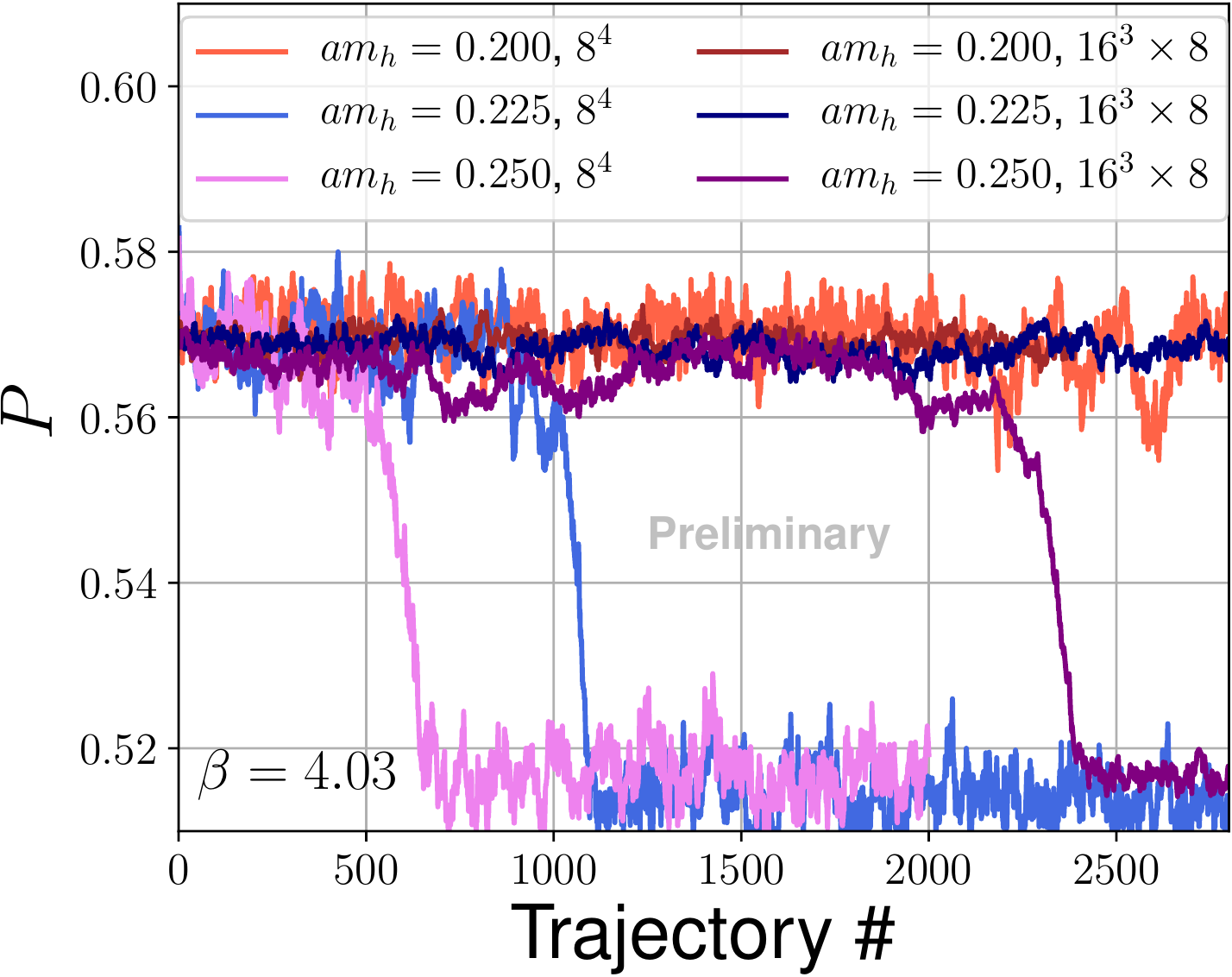}\hspace{.1cm}\includegraphics[width=.225\textwidth]{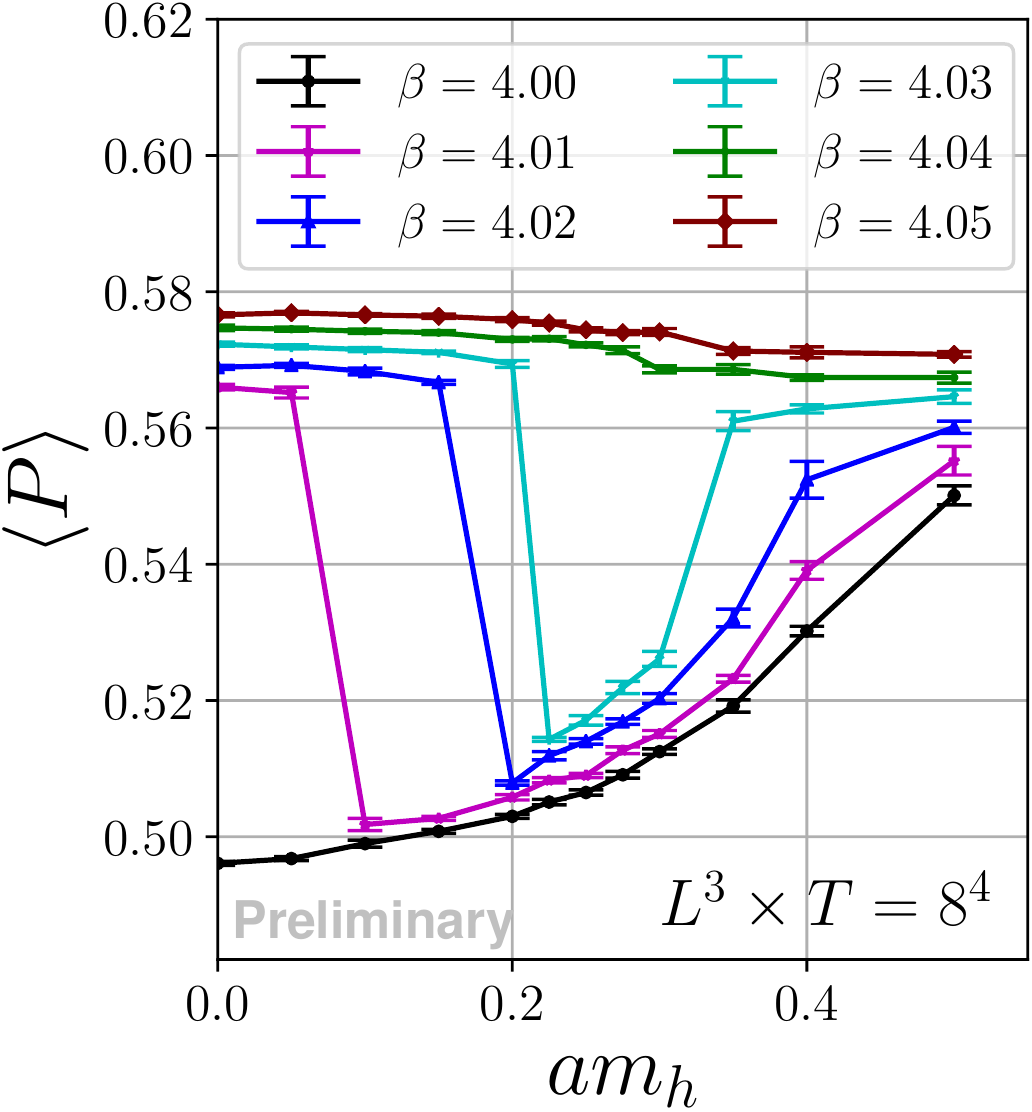}\hspace{.1cm}\includegraphics[width=.255\textwidth]{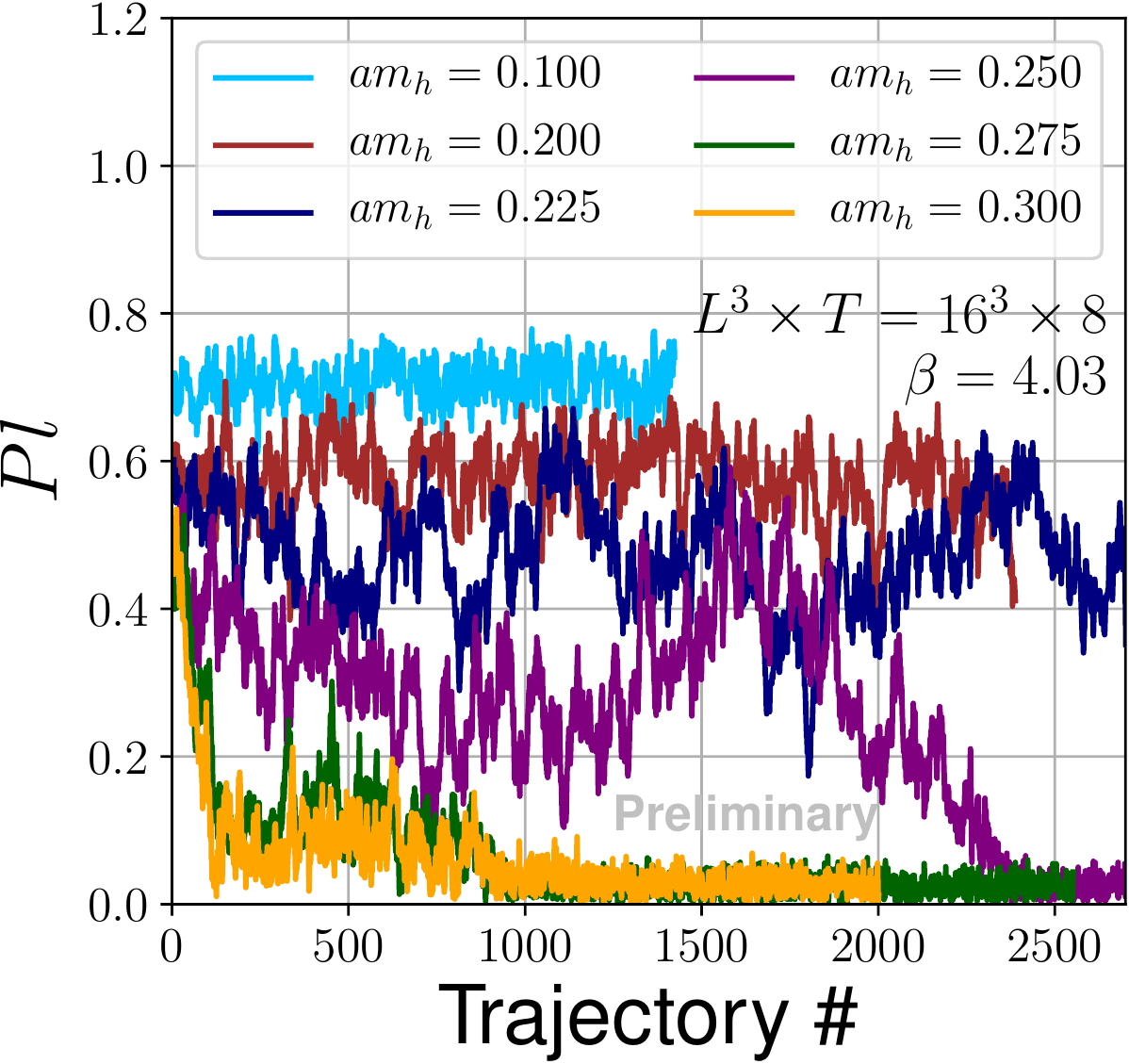}\hspace{.1cm}\includegraphics[width=.185\textwidth]{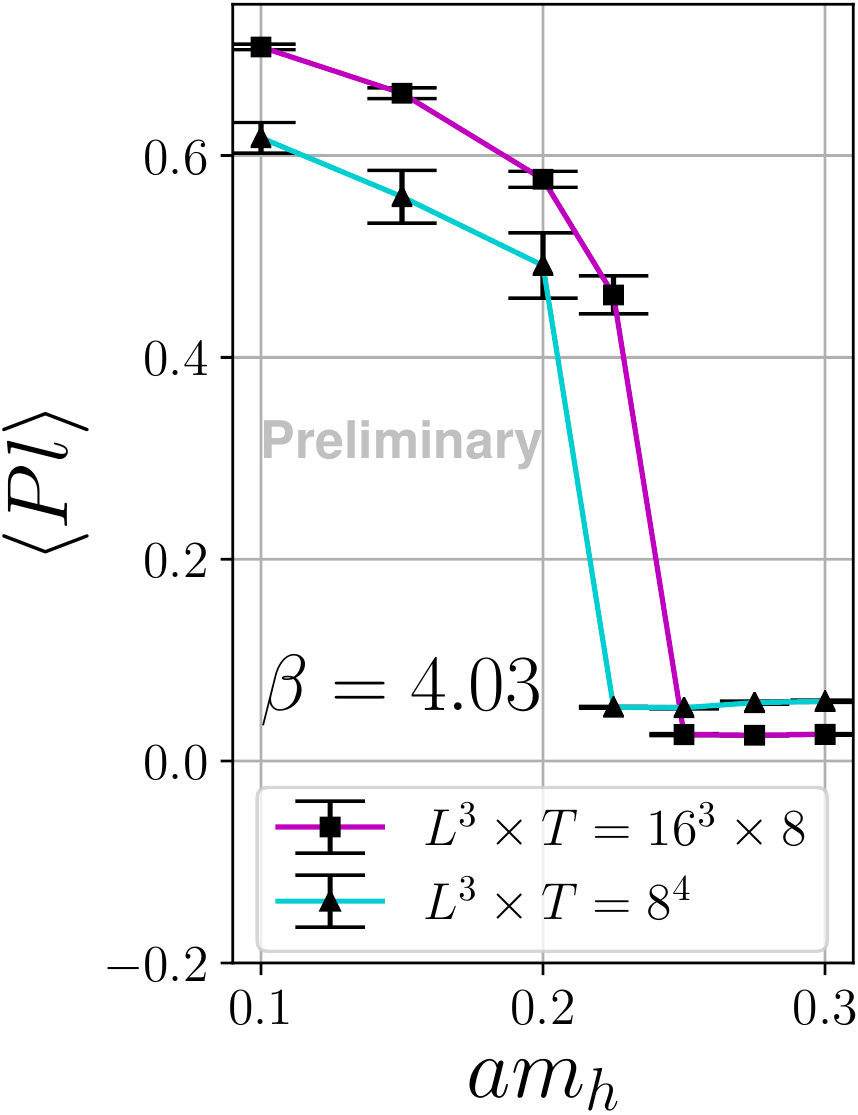}
\caption{Left: Time history of the plaquette $P$ for $8^4$ and $16^3\times 8$ simulations at $\beta=4.03$. Middle left: Ensemble averaged plaquette against $a m_h$
for simulations on $8^4$ lattices. Middle right: Time history of Polyakov loop for $16^3\times 8$ lattices at $\beta=4.03$. Right:
 Comparison of the ensemble averaged Polyakov loop for $8^4$ and $16^3\times 8$ lattices at $\beta=4.03$ for different values of $a m_h$. All
 simulations have $\widehat{m}_\ell/\widehat{m}_h=1$.}
\label{Fig.Finite_T}
\end{figure}

Understanding the phase structure of the $N_f=10$ mass-degenerate and 4+6 mass-split systems
provides insight into their continuum thermodynamic properties and their behavior as lattice models.
On the one hand, the order of the finite-temperature phase transition yields information
about these systems that could be relevant in the early universe. A particularly exciting
possibility is the production of primordial gravitational waves from a first-order
finite-temperature transition that next-generation gravitational wave detectors may unearth
\cite{Brower:2020mab, Caprini:2015zlo, Helmboldt:2019pan, Schettler:2010dp}.
On the other hand, we also expect a zero-temperature bulk phase transition separating a weak
coupling phase with a well-defined continuum limit  from a strong coupling ``bulk'' phase
\cite{Schaich:2012fr, Jin:2012dw, Deuzeman:2012ee}.
The average plaquette $\langle P\rangle$ and average Polyakov loop $\langle Pl\rangle$,
despite not being true order parameters, are sensitive to both transitions \cite{Schaich:2012fr}.
We therefore utilize both observables to understand the location and order of each phase
transition. In Fig.~\ref{Fig.Finite_T}, we present preliminary results for the phase transitions
in the $N_f=10$ mass-degenerate system using $16^3\times 8$ and $8^4$ lattices, where the fourth direction ($aL_4$) corresponds to the inverse temperature.

We map out the bulk phase transition on zero-temperature $8^4$ lattices by keeping the bare
gauge coupling $\beta$ fixed and varying $a m_h$. The left most panel in Fig.~\ref{Fig.Finite_T}
shows the time history of the plaquette, the smallest $1\times 1$ Wilson loop, at gauge coupling  $\beta=4.03$. We observe
that systems with $am_h=0.225$ and 0.250 on $8^4$ volumes ($am_h=0.250$ on $16^3\times 8$) change its phase
after several hundred trajectories in our Markov chain Monte Carlo simulation. The trajectory
number of the transition depends on initial conditions and has no relevance. Calculating ensemble
averages after thermalization effects or observed transitions, we show for the $8^4$ lattices
in the second panel from the left, the ensemble averaged plaquette $\langle P \rangle$ as function of $am_h$ for
several values of the bare coupling $\beta$. These data reveal a first order phase transition in
$a m_h$ from a weak to a strong coupling phase for $4.01 \lesssim \beta\lesssim 4.03$.
The bulk transition weakens and turns into a crossover for $\beta \gtrsim 4.04$. Presently we
are performing $16^4$ simulations to verify that we indeed observe a bulk phase transition.

In addition we study the Polyakov loop to explore the finite temperature phase structure. The third panel from the left shows the Monte Carlo time histories of the Polyakov loop on $16^3 \times 8$ lattices at $\beta=4.03$. Obtaining the ensemble averages for simulations on $16^3\times 8$ and $8^4$ lattices at $\beta=4.03$, we find the finite temperature structure in the fourth panel. Near $am_h\sim 0.2$ the averaged Polyakov loop for both $8^4$ and $16^3\times 8$ lattices sharply drops to zero. There is however a small shift in the value of $am_h$ where this drop occurs. This requires further scrutiny. To better disentangle the finite temperature from the bulk phase transition, simulations on larger volumes with larger $L_4$ are required. While bulk phase transitions do not depend on the volume, finite temperature transitions will move with $L_4$  and hence separate.

In the future, we also plan to measure other observables, such as the chiral condensate, alongside zero- and finite-temperature simulations on larger lattices for a range of different
bare gauge couplings. We are complementing these investigations studying the phase structure of mass-split systems exploring different values of the ratio of $\widehat{m}_\ell/\widehat{m}_h$.

\section*{Acknowledgments}\vspace{-2mm}
We are very grateful to Peter Boyle, Guido Cossu, Antonin Portelli, and Azusa Yamaguchi for developing the \texttt{Grid} software library\footnote{\url{https://github.com/paboyle/Grid}} \cite{Boyle:2015tjk} which provides the basis of this work and thank Andrew Pochinsky and Sergey Syritsyn for developing \texttt{Qlua}\footnote{\url{https://usqcd.lns.mit.edu/w/index.php/QLUA}} \cite{Pochinsky:2008zz} used for measurements. A.H.~and O.W.~acknowledge support by DOE Award No.~DE-SC0010005. This material is based upon work supported by the National Science Foundation Graduate Research Fellowship Program under Grant No.~DGE 1650115. We thank the Lawrence Livermore National Laboratory (LLNL) Multiprogrammatic and Institutional Computing program for Grand Challenge supercomputing allocations. We also thank Argonne Leadership Computing Facility for allocations through the INCITE program.  ALCF is supported by DOE contract DE-AC02-06CH11357. Computations for this work were carried out in part on facilities of the USQCD Collaboration, which are funded by the Office of Science of the U.S.~Department of Energy, the RMACC Summit supercomputer \cite{UCsummit}, which is supported by the National Science Foundation (awards ACI-1532235 and ACI-1532236), the University of Colorado Boulder, and Colorado State University and on Boston University computers at the MGHPCC, in part funded by the National Science Foundation (award OCI-1229059).  We thank ANL, BNL, Fermilab, Jefferson Lab, MGHPCC, LLNL, the NSF, the University of Colorado Boulder, and the U.S.~DOE for providing the facilities essential for the completion of this work.

{\small
  \bibliography{BSM}
  \bibliographystyle{JHEP-notitle}
}

%\begin{thebibliography}{99}
%\bibitem{...}
%....
%\end{thebibliography}

\end{document}